\documentclass[12pt]{article}
\usepackage{amsmath,amssymb,amsfonts,epsf}
\usepackage[margin=1in]{geometry}

\usepackage{fleqn}
\def\be{\begin{equation}}
\def\ee{\end{equation}}
\begin{document}
\begin{center}\ \\ \vspace{60pt}
{\Large {\bf $J/\psi$ and
$\Upsilon$ at high temperature}}\\
\vspace{30pt} I.M.Narodetskii, Yu.A.Simonov, and A.I.Veselov
\vspace{20pt}



\vspace{10pt}
{\it Institute of Theoretical and Experimental Physics, Moscow 117218, Russia}\\

\end{center}

\vspace{30pt}

\centerline{\bf Abstract} \vspace{2mm}

 \noindent  We use the screened Coulomb potential with $r$-dependent coupling constant
  and the non--perturbative quark--antiquark potential
derived within  the Field Correlator Method (FCM) to calculate
$J/\psi$ and $\Upsilon$ binding energies and melting temperatures
in the deconfined phase of quark-gluon plasma.

\vspace{10pt}

\noindent {\bf keywords:}\ \ Quark-gluon plasma, non--perturbative
potential, heavy mesons

\thispagestyle{empty}
\vspace{10pt}

\section{INTRODUCTION}

Since 1986, the 'gold-plated' signature of deconfinement was
thought to be $J/\psi$ suppression \cite{MS:1986}. If Debye
screening of the Coulomb potential above $T_c$ is strong enough
then $J/\psi$ production in A+A collisions will be suppressed.
Indeed, applying the Bargmann condition \cite{bargmann} for the
screened Coulomb potential
\begin{equation}\label{eq:coulomb}
V_C(r)=-\frac{4}{3}\cdot\frac{\alpha_s(r)}{r} \cdot
e^{-m_d\,r},\end{equation} where  $m_d$ is the Debye mass, we
obtain the simple estimate for the number of, say, $c{\overline
c}$ $S$--wave bound states \be n\leq
\mu_c\int\limits_0^{\infty}|V_C(r)|r\,dr\,=\,\frac{4\alpha_s}{3}\cdot\frac{\mu_c}{m_d}\,
,\end{equation} where $\mu_c$ is the constituent mass of the
$c$-quark and for the moment we neglect the $r$-dependence of
$\alpha_s$. Taking $\mu_c=1.4$ GeV and $\alpha_s=0.39$ we conclude
that if $m_d\geq 0.7$ GeV, there is no $J/\psi$ bound state.
Parenthetically, we note that no light or strange mesons ($\mu\sim
300-500$ MeV) survive. But this is not the full story.

There is a significant change of views on physical properties and
underlying dynamics of quark--gluon plasma (QGP), produced at
RHIC, see {\it e.g.} \cite{T:2009} and references there in.
Instead of behaving like a gas of free quasiparticles~--~quarks
and gluons, the matter created in RHIC interacts much more
strongly than originally expected. Also, the interaction deduced
from lattice studies is strong enough to support $Q\overline{Q}$
bound states. It is more appropriate to describe the
non-perturbative (NP) properties of the QCD phase close to $T_c$
in terms remnants of the non--perturbative part of the QCD force
rather than a strongly coupled Coulomb force.

In the QCD vacuum, the NP quark-antiquark potential is {$V=\sigma
r$.} At $T \geq T_c$, $\sigma=0$, but that does not mean that the
NP potential disappears. In a recent paper \cite{NSV:2009} we
calculated binding energies for the lowest $Q\overline{Q}$ and
$QQQ$ eigenstates ($Q=c,b$) above $T_c$ using the NP
$Q\overline{Q}$ potential derived in the Field Correlator Method
(FCM) \cite{NST:2009} and the screened Coulomb potential with the
strong coupling constant $\alpha_s=0.35$ in Eq.
(\ref{eq:coulomb}). In this talk we extend our analysis to the
case of the running $\alpha_s(r)$.

\section{The Field Correlator Method as applied to finite T}  The NP $Q\overline Q$ potential can be studied through the
modification of the correlator functions, which define the
quadratic field correlators of the nonperturbative vaccuum fields
\begin{equation}<tr\,F_{\mu\nu}(x)\Phi(x,0)F_{\lambda\sigma}(0)>\,=\,{\cal
A}_{\mu\nu;\lambda\sigma}\,D(x)+{\cal
B}_{\mu\nu;\lambda\sigma}\,D_1(x),\nonumber\end{equation} where
${\cal A}_{\mu\nu;\lambda\sigma}$ and ${\cal
B}_{\mu\nu;\lambda\sigma}$ are the two covariant tensors
constructed from  $g_{\mu\nu}$ and $x_{\mu}x_{\nu}$, $\Phi(x,0)$
is the Schwinger parallel transporter,
$x$ Euclidian.

At $T\geq\,T_c$, one should distinguish  the color electric
correlators $D^E(x),\,\,\,D^E_1(x)$ and color magnetic correlators
$D^H(x),\,\,\,D^H_1(x)$. Above $T_c$, the color electric
correlator $D^E(x)$ that defines the string tension at $T=0$
becomes zero \cite{Si:1991} and, correspondingly, $\sigma^E=0$.
The color magnetic correlators $D^H(x)$ and $D^H_1(x)$ do not
produce static quark--antiquark potentials, they only define the
spatial string tension $\sigma_s=\sigma^H$ and the Debye mass
$m_d\propto\sqrt{\sigma_s}$ that grows with $T$.

The main source of the NP static $Q\overline Q$ potential at
$T\,\geq\,T_c$ originates from the color--electric correlator
function $D^{E}_1(x)$\begin{equation}\label{eq:potential}
V_{np}(r,T)\,=\,\int\limits_0^{1/T}d\nu(1-\nu T)\int\limits_0^r
\lambda d\lambda\, D_1^{E}(x).\end{equation} In the confinement
region the function $D_1^{E}(x)$ was calculated in \cite{Si:2005}
\begin{equation}D^{E}_1(x)\,=\,{B}\,\,\frac{\exp(-M_0\,x)}{x},\end{equation}
where $ B=6\alpha_s^f\sigma_fM_0$, $\alpha_s^f$ being the freezing
value of the strong coupling constant to be specified later,
$\sigma_f$ is the sting tension at $T=0$, and the parameter $M_0$
has the meaning of the gluelump mass. In what follows we take
$\sigma_f=0.18$ GeV$^2$ and $M_0=1$ GeV. Above $T_c$ the
analytical form of $D_1^E$ should stay unchanged at least up to
$T\sim 2\,T_c$. The only change is $B\to B(T)=\xi(T)B$, where the
factor
$\xi(T)\,=\,\left(1-0.36\frac{M_0}{B}\,\frac{T-T_c}{T_c}\right)$
is determined by lattice data \cite{DMSV:2007}.
Integrating over $\lambda$ one obtains\begin{equation}
V_{np}(r,T)=\frac{B(T)}{M_0} \int\limits_0^{1/T}\, (1-\nu
T)\left(e^{-\nu
M_0}-e^{-\sqrt{\nu^2+r^2}}\,\,M_0\right)d\nu\,=\,V(\infty,T)-V(r,T)\end{equation}
Note that $V(\infty,T_c)\approx 0.5$ GeV that agrees with lattice
estimate for the free quark-antiquark energy.

In the framework of the FCM, the masses of heavy quarkonia are
defined as \begin{equation} \label{eq:mass}M_{Q\bar
Q}\,=\,\frac{m_Q^2}{\mu_{Q}}\,+\,\mu_Q\,+\,E_0(m_Q,\mu_Q),\end{equation}
$E_0(m_Q,\mu_Q)$ is an eigenvalue of the Hamiltonian
$H\,=\,H_0\,+\,V_{np}\,+\,V_C$, $m_Q$ are the bare quark masses,
$\mu_Q$ are the auxiliary fields that are introduced to simplify
the treatment of relativistic kinematics. The auxiliary fields are
treated as c--number variational parameters to be found from the
extremum condition imposed on $M_{Q\bar Q}$ in Eq.
(\ref{eq:mass}).
Such an approach allows for a very transparent interpretation of
axiliary fields as the constituent masses that appear due to the
interaction. Once $m_Q$ is fixed, the quarkonia spectrum is
described.
The dissociation
points are defined as those temperature values for which the
energy gap between $V(\infty,T)$ and $E_0$ disappears.
\section{Coulomb potential} We use the perturbative screened
Coulomb potential (\ref{eq:coulomb}) with the $r$-dependent QCD
coupling constant $\alpha_s(r,T)$. Note that in the entire regime
of distances for which at $T=0$ the heavy quark potential can be
described well by QCD perturbation theory $\alpha_s(r,T)$ remains
unaffected by temperature effects at least up to $T\leqslant
3\,T_c$ and agrees with the zero temperature running coupling
$\alpha_s(r,0)=\alpha_s(r)$. For our purposes, we find it
convenient to define the $r$--dependent coupling constant in terms
of the ${\textbf q}^2$ dependent constant $\alpha_B({\textbf
q}^2)$ calculated in the background perturbation theory (BPTh)
\cite{Si:1995}

\begin{equation}
\label{eq:alpha}
\alpha_s(r)\,=\,\frac{2}{\pi}\,\int\limits_0^{\infty}dq\,\,\frac{\sin\,qr}{q}\,\,\alpha_B({\textbf
q}^2).
\end{equation}
The formula for $\alpha_B({\textbf q}^2)$ is obtained by solving
the two-loop renormalization group equation for the running
coupling constant in QCD:
\begin{equation}
\label{eq:alpha_V}
 \alpha_B({\textbf
q}^2)\,=\,\frac{4\pi}{\beta_0\,t}\left(1\,-\,\frac{\beta_1}{\beta_0\,^2}\,\frac{\ln
t}{t}\right),\,\,\,\,\,t\,=\,\ln\,\frac{{\textbf
q}^2+m_B^2}{\Lambda_V^2},
\end{equation}
where $\beta_i$ are the coefficients of the QCD $\beta$-function.

In Eq. (\ref{eq:alpha_V}) the parameter $m_B\sim\,1$ Gev has the
meaning of the mass of the lowest hybrid excitation. The result
can be viewed as arising from the interaction of a gluon with
background vacuum fields.  We employ the values
$\Lambda_V\,=\,0.36\,{\rm GeV},\,\,\, m_B\,=\,0.95\,{\rm GeV}$,
which lie within the range determined in Ref. \cite{BK2001}. The
result is consistent with the freezing of $\alpha_B(r)$ with a
magnitude 0.563 (see Table 4 of Ref. \cite{KNV:2009}. The zero
temperature potential with the above choice of the parameters
gives a fairly good description of the quarkonium spectrum
\cite{BK2001}. At finite temperature we utilize the information on
$m_d$ in Eq. (\ref{eq:coulomb}) from Ref. \cite{A:2003}. For
pure-gauge SU(3) theory ($T_c=275$ MeV) $m_d$ varies between 0.8
GeV and 1.4 GeV, when $T$ varies between $T_c$ and $2\,T_c$.

\section{Results}
 \begin{table} \caption{$J/\psi$
above the deconfinement region.  $V(\infty,T)$ is the continuum
threshold (a constant shift in the potential). Units are GeV or
GeV$^{-1}$.}\vspace{2mm}

\centering
\begin{tabular}{cccccc}\\\hline\\
$T/T_c$&~~$V(\infty)$&$\mu_b$&$E_0-V(\infty)$&$r_0$&$M_{J/\psi}$\\\\
1&~~0.445&~~1.443&-0.011&8.23&3.235\\
1.2~~&~~0.368&~~1.423&-\,0.003&10.07&3.171\\
\\\hline

\end{tabular}

 \label{tab:cc}\end{table}

The solutions for the binding energy for the $1S$ $J/\psi$ and
$\Upsilon$ states are shown in Tables \ref{tab:cc}, \ref{tab:bb}.
In these Tables we present the constituent quark masses $\mu_Q$
for $c\overline{c}$ and $b\overline{b}$, the differences
$\varepsilon_Q=E_0\,-\,V_{Q\overline{Q}}(\infty)$, the mean
squared radii $r_0\,=\,\sqrt{\,\overline{r^2}\,}$, and the masses
of the $Q\overline{Q}$ mesons. We employ $m_c=1.4$ GeV and
$m_b=4.8$ GeV. Note that, as in the confinement region, the
constituent masses $\mu_Q$ only slightly exceed bare quark masses
$m_Q$ that reflect smallness of the kinetic energies of heavy
quarks.

At $T=T_c$ we obtain the weakly bound $c{\overline c}$ state that
disappears  at $T\sim 1.3\,T_c$. The charmonium masses lie in the
interval 3.2 - 3.3 GeV, that agrees with the results of Ref.
\cite{DMSV:2007}. Note that immediately above $T_c$ the mass of
the  $c\overline{c}$ state is about 0.2 GeV higher than that of
$J/\psi$. As expected, the $\Upsilon$ state remains intact up to
the larger temperatures, $T\,\sim\, 2.3\,T_c$, see Table
\ref{tab:bb}. The masses of the L = 0 bottomonium  lie in the
interval 9.7--9.8 GeV, about 0.2--0.3 GeV higher than 9.460 GeV,
the mass of $\Upsilon(1S)$ at $T=0$. At $T=T_c$ the $b{\overline
b}$ separation $r_0$ is 0.25 fm that compatible with $r_0=0.28$ fm
at $T=0$. The 1S bottomoniium undergo very little modification
till
 $T\,\sim 2\,Tc$.
The results agree with those found previously for a constant
$\alpha_s=0.35$ \cite{NSV:2009}. We also mention that the melting
temperatures for $\Omega_c$ and $\Omega_b$ calculated in
\cite{NSV:2009} practically coincide with those for $J/\psi$ and
$\Upsilon$.

Our results for $1S(J/\psi)$ are qualitatively agree with those of
Refs. \cite{Blaschke:2005}, \cite{Alberico} based on
phenomenological $Q\overline{Q}$ potentials identified with the
free quark-antiquark energy measured on the lattice while our
melting temperature for $1S(\Upsilon)$ is much smaller than $T\sim
(4-6)\,T_c$ found in Ref. \cite{Alberico}.

\begin{table}
\caption{$1S\,\,b\overline b$ state above the deconfinement
region.
}\vspace{4mm} \centering
\begin{tabular}{c|ccccc}\hline\\
$T/T_c$&~~{ $V(\infty,T)$}&$\mu_b$&{ $E_0(T)-V
(\infty,T)$}&$r_0$&$M_{b\overline{b}}$\\\\
1&~~{ 0.445}&~~4.948&{ -0.255}&1.39&9.796\\
1.3&~~{ 0.332}&~~4.922&{ -0.15}8&1.69&9.777\\
1.6&~~{ 0.237}&~~4.894&{ -\,0.084}&2.23&9.755\\
2.0&~~{ 0.134}&~~4.854&{ -\,0.022}&4.23&9.712\\
2.2&~~{ 0.090}&~~4.831&{ -\,0.006}&6.77&9.684\\
2.3&~~{ 0.070}&~~4.821&{-\,0.002}&8.32&9.668

\\ \\\hline
\end{tabular}
 \vspace{3mm}

 \label{tab:bb}\end{table}

 This work was supported in part by RFBR Grants
$\#\,\,$08-02-00657, $\#\,\,$08-02-00677, $\#\,\,$09-02-00629 and
by the grant for scientific schools $\#\,\,$NSh.4961.2008.2.

\end{document}